\documentclass[prb,twocolumn,amsmath,superscriptaddress,amssymb,longbibliography]{revtex4-2}

\usepackage{graphicx}
\usepackage{dcolumn}
\usepackage{bm}
\usepackage{amssymb}
\usepackage{amsmath}
\usepackage{wasysym}
\usepackage{color}
\usepackage{times}

\usepackage{placeins}
\usepackage{epstopdf}
\usepackage{array}

\usepackage[separate-uncertainty=true, multi-part-units=single]{siunitx}

\usepackage{ulem}
\usepackage{nicefrac}
\usepackage[dvipsnames]{xcolor}

\usepackage[colorlinks]{hyperref} 
\hypersetup{breaklinks=true, citecolor=blue, linkcolor=blue, menucolor=blue, urlcolor=blue}

\usepackage{comment}
\usepackage{textcase}

\usepackage[capitalise]{cleveref}

\newcommand{\FGT}{Fe${_3}$GeTe$_{2}$}

\makeatletter
\renewcommand*\env@matrix[1][c]{\hskip -\arraycolsep
  \let\@ifnextchar\new@ifnextchar
  \array{*\c@MaxMatrixCols #1}}
\makeatother
\begin{document}

\title{Growth of Altermagnetic $\alpha$-MnTe Films: Substrate Variation and Surface Modification}

\author{M. Dittmar}
\author{L. Hirnet}
\author{H. Haberkamm}
\author{F. Beisler}
\author{C.-W. Chuang}
\author{R. Ganser}
\author{P. Kagerer}
\affiliation{Experimentelle Physik VII and W\"{u}rzburg-Dresden Cluster of Excellence ctd.qmat, Universit\"{a}t W\"{u}rzburg, Am Hubland, D-97074 W\"{u}rzburg, Germany}

\author{M.-J. Huang}
\affiliation{
	Ruprecht Haensel Laboratory, Deutsches Elektronen-Synchrotron DESY, D-22607 Hamburg, Germany}

\author{J. Buck}
\affiliation{
	Ruprecht Haensel Laboratory, Deutsches Elektronen-Synchrotron DESY, D-22607 Hamburg, Germany}
\affiliation{
	Institut f\"ur Experimentelle und Angewandte Physik, Christian-Albrechts-Universit\"at zu Kiel, D-24098 Kiel, Germany}

\author{M. Hoesch}
\affiliation{
	Deutsches Elektronen-Synchrotron DESY, D-22607 Hamburg, Germany}

\author{K. Rossnagel}
\affiliation{
	Ruprecht Haensel Laboratory, Deutsches Elektronen-Synchrotron DESY, D-22607 Hamburg, Germany}
\affiliation{
	Institut f\"ur Experimentelle und Angewandte Physik, Christian-Albrechts-Universit\"at zu Kiel, D-24098 Kiel, Germany}

\author{M. \"{U}nzelmann}
\email{maximilian.uenzelmann@uni-wuerzburg.de}
\affiliation{Experimentelle Physik VII and W\"{u}rzburg-Dresden Cluster of Excellence ctd.qmat, Universit\"{a}t W\"{u}rzburg, Am Hubland, D-97074 W\"{u}rzburg, Germany}

\author{F. Reinert}
\affiliation{Experimentelle Physik VII and W\"{u}rzburg-Dresden Cluster of Excellence ctd.qmat, Universit\"{a}t W\"{u}rzburg, Am Hubland, D-97074 W\"{u}rzburg, Germany}

\date{\today}

\begin{abstract}
Manganese telluride (MnTe) in its hexagonal $\alpha$-MnTe crystal structure has evolved as one of the altermagnet workhorse materials. The synthesis of MnTe thin films is highly relevant for both fundamental science and device applications. 
Here, we report on the epitaxial growth of MnTe thin films and heterostructures. The films are studied by X-ray and electron diffraction as well as soft X-ray angle-resolved photoemission spectroscopy.
We demonstrate the ability to grow high-quality $\alpha$-MnTe on various substrates, ranging from transparent band insulators, over topological insulators, metallic transition metal chalcogenides, to the van der Waals ferromagnet \FGT. While insulating substrates are useful for transport experiments or optical spectroscopy, metallic topological surface states may trigger spintronic interface effects, such as spin-orbit torques. Metallic substrates, in general, are highly relevant to avoid charging at the insulating MnTe films in electron spectroscopy or microscopy methods. Lastly, ferromagnetic substrates will be of interest to control magnetization across the interface. In addition, we discuss the formation of superstructures on the MnTe(0001) surfaces, that emerge directly after growth, upon subsequent tellurium evaporation and after thermal treatment. 
This will be relevant in further studying the surface magnetic and electronic structure in $\alpha$-MnTe.
\end{abstract}

\maketitle

\section{Introduction}

\begin{figure*}
    \centering
    \includegraphics[width=0.9\linewidth]{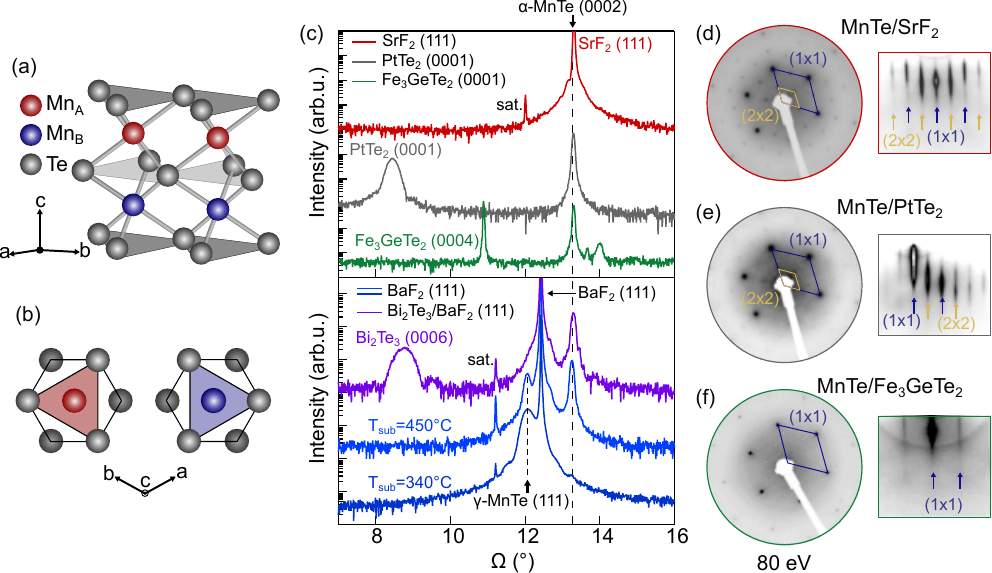}
    \caption{(a),(b) Crystal structure of $\alpha$-MnTe with the opposite Mn sublattices marked in red and blue. The differently shaded Te atoms in (b) represent the different environments of both Mn sublattices, which are connected by a sixfold screw axis symmetry. (c) XRD $\omega-2\Theta$ measurements of epitaxial MnTe thin films grown on different substrates. For the BaF$_2$ substrate measurements of two films grown at different substrate temperatures are shown. The dashed vertical lines indicate the expected positions of the (111) peak of the cubic MnTe phase and the (0002) of the hexagonal $\alpha$-MnTe phase. The respective substrate peaks are  labeled and satellite lines are indicated as 'sat'. (d-f) RHEED and LEED images of the MnTe films grown on SrF$_2$ and PtTe$_2$ and \FGT, respectively. Streaky RHEED patterns and sharp LEED spots indicate flat surfaces with large domains in all cases.} 
    \label{fig:1}
\end{figure*}

The theoretical prediction of altermagnetism (AM) has generated great interest in condensed matter physics \cite{PhysRevX.12.031042, PhysRevX.12.040501,Jungwirth2026}. Among the suggested altermagnetic material candidates, manganese telluride (MnTe) in its hexagonal crystal structure --- referred to as $\alpha$-MnTe (Fig.~\ref{fig:1} (a,b)) --- has turned out to be one of the most promising workhorse materials \cite{Mazin2023, Lovesey2023}. Experimental evidence of AM has been obtained, for example, from the anomalous Hall effect \cite{Gonzales2023, Kriegner2016, smolenski2025}, X-ray magnetic circular dichroism \cite{Hariki2024,Amin2024}, and angle-resolved photoemission spectroscopy (ARPES) \cite{Krempasky2024, Lee2024, Osumi2024, Hajlaoui2024}.

The growth of thin films plays a decisive role in both fundamental research and application-related aspects. For $\alpha$-MnTe, particularly SrF$_2 (111)$ and InP$(111)$ have proven to be suitable substrates that allow for the growth of high-quality films via molecular beam epitaxy (MBE) \cite{Kriegner2017}.
Recent studies have demonstrated the crucial influence of strain relaxation via misfit arrays in $\alpha$-MnTe films grown on GaAs$(111)$ \cite{Bey2025}, where the anomalous Hall effect shows an unexpected tunability \cite{Bey2024}.
Overall, the tremendous increase in experimental efforts on $\alpha$-MnTe demands further growth studies and the availability of 'straightforward' growth recipes that ensure the reproducibility and comparability of experimental results. 
Moreover, broadening the substrate spectrum is essential for accessing novel interface phenomena and functionalities. For example, substrates with strong spin–orbit coupling may promote spin–orbit torque generation via the spin Hall effect, while ferromagnetic substrates could enable magnetic coupling across the interface and potentially modify the magnetic domain structure in the MnTe film.
In addition to this, surface effects on altermagnets \cite{zhao2026,zhou2026} require further insights into surface properties and ways to modify the surface atomic structure.

Here, we focus on these aspects and study the MBE growth of $\alpha$-MnTe on various substrates, ranging from transparent wide-gap band insulators to topological and ferromagnetic van der Waals (vdW) materials (see Table~\ref{tab_sub}). The epitaxial films are investigated by X-ray and electron diffraction techniques as well as photoemission spectroscopy.
We compare growth properties for the different substrates used and particularly shed light on the interplay between lattice mismatch, interface chemistry, and growth temperature. Furthermore, we systematically investigate the formation of different surface superstructures during growth, upon post-growth tellurium exposure, and after subsequent thermal treatment. 
Finally, our soft X-ray (SX) ARPES data confirm a spin-split electronic band structure, underpinning the high quality of the grown films.

\section{Experimental}

The MnTe thin films were grown in ultra-high vacuum (UHV) by MBE, in a MBE system described in earlier works \cite{Kagerer2020}. We used thermal effusion cells charged with elemental manganese and tellurium. The beam equivalent pressures (BEP) of the elemental fluxes were measured with a retractable Bayard-Alpert type ionization gauge. For all substrates discussed here, the optimal flux ratio was found to be around BEP$_\text{Te}$/BEP$_\text{Mn}$\,=\,7.
The thickness of all films shown in this work is above $60\,\mathrm{nm}$.\\
\textit{In-situ} reflection high-energy electron diffraction (RHEED) measurements were done with a Dr. Gassler RHEED system at substrate temperatures ranging from room temperature to the growth temperature. Low-energy electron diffraction (LEED) measurements were carried out \textit{in-situ} after growth using an OCI LEED 800 spectrometer, with the samples being at room temperature during the measurements. \textit{Ex-situ} XRD measurements were done at room temperature with a Bruker D8-discover diffractometer \cite{Kagerer2020}.\\
Photoemission experiments were done at the ASPHERE III endstation at the P04 beamline at DESY Hamburg \cite{Figgemeier2025}. For this, samples were prepared in Würzburg and transported to Hamburg using a UHV suitcase. The measurements were then carried out at a temperature of $30\,$K with a nominal energy resolution $< 120\,\mathrm{meV}$ for all measurements.

\begin{table}[h]
    \centering
        \caption{Substrates used in this work for the growth of $\alpha$-MnTe, including the corresponding lattice mismatch and physical substrate properties.}
    \begin{tabular}{ l c c r}
        \hline
        \hline
        \vspace{0.2cm}
        substrate & lattice mismatch  &properties  \\
         \hline
       SrF$_2$ (111)  & 1.2\,\% & transparent, insulating  \\

       BaF$_2$ (111)  & -5.3\,\% & transparent, insulating \\

        Bi$_2$Te$_3$ (0001) & -5.3\,\% & vdW topological insulator \\

         PtTe$_2$ (0001) & 4.0\,\%& vdW topological metal  \\
         
         \FGT (0001) & 4.0\,\% & vdW ferromagnet, metallic \\
         \hline
         \hline
    \end{tabular}
    \label{tab_sub}
\end{table}

\section{Results and Discussion}

\subsection{Epitaxial growth of MnTe on different substrates}

\subsubsection{Growth on SrF$_2$, BaF$_2$, and Bi$_2$Te$_3$~/~BaF$_2$}

We first recall the epitaxial growth of $\alpha$-MnTe on SrF$_2$(111), which can be considered as the ideal substrate, given the small lattice mismatch (see Table~\ref{tab_sub}) and matching coefficient of thermal expansion \cite{Kommichau1986,Minikayv2015}.
In the XRD spectrum (Fig.~\ref{fig:1}(c)), we find the $\alpha$-MnTe $(0002)$ peak close to the substrate feature, indicating the growth of the desired hexagonal $\alpha$-phase, as expected from previous studies \cite{Kriegner2017}. This holds for all substrate temperatures $T_\mathrm{sub}$ used, with the best film quality being achieved at $T_\mathrm{sub}=440\,^\circ$C. The RHEED and LEED patterns in Fig.~\ref{fig:1}(d) show sharp diffraction streaks and spots, respectively, indicating a layer-by-layer growth with long-range periodic order. Diffraction spots highlighted by dark blue arrows and rhombus in Fig.~\ref{fig:1}(d) correspond to the expected $\alpha$-MnTe$(0001)-(1\times 1)$ (i.e., bulk-terminated) surface unit cell. Moreover, as will be discussed in more detail in Section~\ref{Sect_Surface}, the appearance of half-order spots (yellow) proves the formation of a  $(2\times 2)$ surface reconstruction.
We point out that we observe this superstructure for various films (i.e., grown on all substrates except \FGT), which crystallize in the desired $\alpha$-phase. As such, the $(2\times 2)$ superstructure in many cases serves as a first indirect \textit{in-situ} fingerprint of a 'successful' film growth.

In order to investigate the influence of lattice mismatch, we have grown MnTe films on BaF$_2(111)$ substrates. BaF$_2$ is structurally similar to SrF$_2$, but has a larger lattice constant with a lattice mismatch of $\approx -5.3\,$\% to bulk $\alpha$-MnTe.
We first consider the growth directly on the uncovered BaF$_2(111)$ surface. The corresponding XRD spectra are depicted in light and dark blue in Fig.~\ref{fig:1}(c) for films grown at substrate temperatures of $T_\mathrm{sub}=450\,^\circ$C and $T_\mathrm{sub}=340\,^\circ$C, respectively.
In contrast to SrF$_2$, in which we find the substrate temperature only affects the film quality but not the phase in which the films grow, we here observe a decisive influence of $T_\mathrm{sub}$: At reduced temperatures, we observe a Bragg peak (Fig.~\ref{fig:1}(c)) at smaller angles, i.e., slightly below the (111) bulk BaF$_2$ feature and in good agreement with  the cubic $\gamma$-MnTe phase. 
If we increase the substrate growth temperature, we still see the $\gamma$-MnTe peak but in this case, the desired Bragg peak of the $\alpha$-phase appears as well, suggesting that both phases are present in the film.
We interpret this such that the interface between film and BaF$_2(111)$ substrate favors forming the $\gamma$-MnTe phase, because of the better lattice matching, as compared to $\alpha$-MnTe. Towards the vacuum, however, the latter seems to be favored for higher growth temperatures. This means, in the latter case, the film starts growing in the cubic phase to form a smooth interface, but transforms into the hexagonal phase at higher layers.
This is further supported by the observation of the $(2\times 2)$-reconstruction, indicative of an $\alpha$-MnTe(0001) surface, in surface-sensitive LEED measurements (see Fig.~S1 in the SM \cite{supp}).

We next study the impact of a modified interface on the growth. That is, we first grow a few 'buffer' layers of Bi$_2$Te$_3$ on BaF$_2$ and subsequently initiate the growth of MnTe.
Bi$_2$Te$_3(0001)$ has an almost ideal lattice match with BaF$_2(111)$, resulting in extremely flat films with long-range order \cite{Fornari2016}. Moreover, Bi$_2$Te$_3$ is an archetype three-dimensional topological insulator \cite{Chen2009} built from vdW layers.
We find that --- even though the lattice mismatch has essentially not changed as compared to the clean BaF$_2(111)$ substrate --- the MnTe film directly grows in the $\alpha$-phase and no $\gamma$-MnTe peak is observed anymore in the corresponding XRD spectrum (see Fig.~\ref{fig:1}(c)). This means that $\alpha$-MnTe can form lattice-mismatched interfaces with a vdW substrate Bi$_2$Te$_3$, but not with BaF$_2$, where the interface bonding likely has a dominant ionic and covalent character.

\subsubsection{Growth on platinum tellurides}

Having established the growth on a vdW substrate with considerable lattice mismatch, we will now focus on the growth of MnTe on another class of vdW materials, namely transition metal chalcogenides (TMCs). In particular, we focus on metallic platinum telluride (PT) compounds PtTe$_2$ and Pt$_3$Te$_4$. These belong to a class of topological metals with spin-momentum-locked surface states \cite{Yan2017,Fujii2021,Qahosh2025}. The associated large spin Hall conductivity makes PT an interesting building block in heterostructures for spin switching devices \cite{Wang2024}.
PtTe$_2$ crystallizes in the trigonal (1T) structure with vDW layers being stacked along the $[0001]$ direction. Pt$_3$Te$_4$ is composed of alternating layers of PtTe$_2$ and Pt$_2$Te$_2$. Both compounds have the same in-plane lattice constant of $a=4.00\,\mathring{\mathrm{A}}$ \cite{Lasek2022}, which results in a lattice missmatch of $\approx 4\,$\% to $\alpha$-MnTe.
In Fig.~\ref{fig:1}(c), we exemplarily show the XRD spectrum of a MnTe film grown on a PtTe$_2(0001)$ substrate, clearly demonstrating the growth of MnTe in the desired hexagonal $\alpha$-phase. Again, we find that this holds for all tested growth temperatures, with the best film quality achieved at $T_\mathrm{sub} \approx 500\,^\circ$C, i.e., slightly higher than that of the SrF$_2$ and BaF$_2$ substrates. Note that we do not find a significant influence of using either Pt$_3$Te$_4$ or PtTe$_2$ substrates on the growth properties, as expected from the structural similarity. 
Rather, the Te-poor Pt$_3$Te$_4$ surface tends to transform into Te-rich PtTe$_2$ during the initial growth of MnTe, because of the high Te flux.
\\
Like in the case of SrF$_2$, the streaky RHEED pattern (Fig.~\ref{fig:1}(e)) demonstrates layer-by-layer growth of ordered films without significant formation of 3D islands. The sharp LEED spots, moreover, indicate the presence of the $(2\times 2)$ reconstruction and a good surface quality with long-range periodic order. However, as compared to SrF$_2$ and BaF$_2$, we find slightly broader LEED spots, indicating an overall smaller mean island size. In Section~\ref{Sect_ARPES}, we return to the influence of structural quality on band-structure measurements via SX-ARPES (see Fig.~\ref{fig:ARPES}(a,b)).
Overall, we have established the growth of $\alpha$-MnTe on the vdW topological metal PtTe$_2 (0001)$.

\begin{figure*}[t!]
    \centering
    \includegraphics[width=0.9\linewidth]{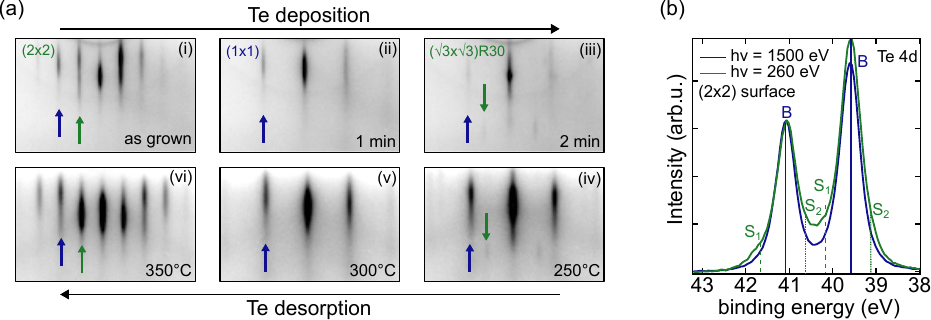}
    \caption{(a) In-situ RHEED patterns illustrating the evolution of surface superstructures on $\alpha$-MnTe$(0001)$~/~SrF$_2$ during Te deposition ((i)–(iii), (BEP = $1.3 \cdot 10^{-7}$ mbar, $T_{\text{sub}} = 80\,^\circ$C) and their subsequent modification through Te desorption induced by post-growth annealing ((iv)–(vi)). The Te deposition durations and annealing temperatures are indicated directly in the corresponding RHEED images. Fundamental MnTe reflections and superstructure-related diffraction features are highlighted by arrows. (b) XPS spectra of the Te 4d core-level region acquired at different photon energies from a MnTe surface exhibiting a $(2 \times 2)$ superstructure.}
    \label{fig:surface_structure}
\end{figure*}

\subsubsection{Growth on the vdW ferromagnet \FGT}
Further extending the range of suitable substrates, $\alpha$-MnTe films are also grown on Fe$_3$GeTe$_2$ --- a prototypical van der Waals magnet with a layered crystal structure that retains ferromagnetism down to the monolayer limit \cite{Gibertini2019}. 
As a strongly anisotropic ferromagnet with an out-of-plane easy axis and Curie temperature around $220–230\,$K (bulk), \FGT~  combines metallic magnetism with structural stability, making it attractive for magnetic heterostructures, in general. The lattice mismatch to $\alpha$-MnTe is $\approx4\%$, i.e., almost identical to PtTe$_2$, rendering it a promising substrate candidate for investigating the growth of MnTe \cite{Liu2017}.
Moreover, as layered vdW substrates, \FGT~single crystals can be easily cleaved in UHV, yielding very good surface quality, ideally suitable for epitaxial growth.\\
In Fig. \ref{fig:1}(c), the XRD curve of a film grown on \FGT~is shown, demonstrating the epitaxial growth of $\alpha$-MnTe on the (0001) surface of Fe$_3$GeTe$_2$, in line with previous studies on epitaxial \FGT~films \cite{Liu2017}. Best film quality is achieved at slightly lower substrate temperatures of about $T_\mathrm{sub}=410\,^\circ$C. Again, we find a streaky RHEED pattern (Fig.~\ref{fig:1}(f)), and the LEED spots are the smallest we could achieve (among all films grown on all substrates), indicating large surface terraces. Surprisingly, $\alpha$-MnTe films on \FGT~ are the only ones, for which we do not observe a $(2\times2)$ superstructure. This holds for all films grown under different growth conditions, as further elaborated on in Section~\ref{Sect_Surface}.

\subsection{Surface superstructures}
\label{Sect_Surface}

We will now focus on the emergence of surface superstructures on our grown $\alpha$-MnTe(0001) films. As mentioned before, all films (except for those grown on \FGT) exhibit a $(2 \times 2)$ surface reconstruction. Representatively, we here focus on $\alpha$-MnTe/SrF$_2$ for the further investigation of this surface.
In Fig.~\ref{fig:surface_structure}(b), we show X-ray photoemission spectra at the Te-4d$_{3/2}$ and 4d$_{5/2}$ core levels. The spectra show a major component, labeled $B$, at energies of 39.58\,eV and 41.08\,eV. We attribute this to the tellurium embedded in the $\alpha$-MnTe bulk crystal environment (see Fig.~\ref{fig:1}(a,b)).
When reducing the photon energy from 1500\,eV to 260\,eV, i.e., increasing the surface sensitivity, additional components appear in the spectrum as shoulders left and right of the main Te-4d components. These features are assigned to two additional doublets (labeled S$_1$ and S$_2$ in Fig. \ref{fig:surface_structure}(b)) arising from Te atoms at the surface. In other words, there are (at least) two types of Te species with a chemical environment that differs from the $\alpha$-MnTe bulk. This, in turn, suggests that the $(2\times 2)$ unit cell contains differently bound Te atoms.
This could be either a $(2 \times 2)$ Te adlayer on top of the bulk-like-terminated $(1\times 1)$ surface or a Te-deficient surface, likewise being found in isostructural MnAs(0001) films \cite{Ouerghi2006}. Given the thermal stability of the $(2 \times 2)$ reconstruction --- up to temperatures of at least $T_\mathrm{sub} = 500\,^\circ$C --- observed here, we expect the former scenario to be more unlikely. That is, rather weakly bound Te adlayers may desorb at lower temperatures, about $300\,^\circ$C (within the range of conventional Te-cell temperatures). However, in the case of a Te-deficient topmost $\alpha$-MnTe$(0001)$ layer, the question arises whether Te vacancies can be removed by additional deposition of Te.

In order to test this and, more broadly, to investigate the influence of post-growth surface treatment and the possibility of altering the MnTe surface structure, we have studied the evolution of surface superstructures upon Te deposition and subsequent post-growth annealing.
This was monitored live by acquiring \textit{in-situ} RHEED patterns shown in Fig. \ref{fig:surface_structure} (a, i-vi). Corresponding LEED pattern are shown in Fig. S2 of the SM \cite{supp}.
First, Te was deposited onto the as-grown $(2\times 2)$ surface (i) for three minutes, in total, with a BEP of $p_\mathrm{Te}=1.3 \cdot 10^{-7}\,$mbar, while keeping the sample at a temperature of $T_\mathrm{sub}=80\,^\circ$C to activate a slight surface diffusion but avoid immediate re-desorption. Upon Te deposition, the intensity of the $(2 \times 2)$ superstructure streaks decreases and, after one minute, only the integer reflections of the unreconstructed $(1 \times 1)$ surface are visible (ii). Further deposition immediately gives rise to new reflections (iii), which can be assigned to a $(\sqrt{3} \times \sqrt{3})R30^\circ$ surface reconstruction. This is further supported by the corresponding LEED pattern shown in Fig. S2 of the SM \cite{supp}. Shortly after step (iii), additional diffraction streaks and spots arise, which can most likely be attributed to the initial formation of bulk-like Te adlayers (see SM Fig. S2 \cite{supp}).
\\
To check whether this process is reversible via Te desorption, we investigated the superstructure evolution upon post-annealing of the grown Te thin film.
At a temperature of $250\,$°C, again only the $(\sqrt{3} \times \sqrt{3})R30^\circ$ superstructure is present (iv). Increasing the temperature, the reflections assigned to this superstructure vanish until only the main reflections are visible in the pattern at $\approx 300\,$°C (v). Further annealing leads to the reappearance of the half-order reflections and therefore the full recovery of the as-grown $(2 \times 2)$ reconstructed surface at $\approx 350\,$°C (vi).

Based on these observations, we interpret the respective superstructures found as follows: The thermally most stable $(2\times 2)$ reconstruction could, in fact, be assigned to the Te-deficient uppermost $\alpha$-MnTe layer. We can only speculate about the exact atomic structure, but one possible scenario is a Te kagome-like layer, suggested in isostructural MnAs(0001) films \cite{Ouerghi2006}. Kagome-type Te surface layers have also been found in the $(3\times 3)$-Te superstructure on Pt(111) \cite{Kisslinger2023}, yielding comparable surface lattice constants, $2a_\mathrm{MnTe(0001)} \approx 3a_\mathrm{Pt(111)} \approx 8.32\,\mathrm{\mathring{A}}$.
The surface vacancies can be filled in a \textit{vaccancy-healing process} upon Te exposure, which results in a bulk-like-terminated $\alpha$-MnTe(0001) surface (Fig.~\ref{fig:surface_structure}(a, ii and v). We have to mention that, based on our efforts, this $(1\times 1)$ structure is not very stable. For temperatures only slightly higher than $300\,$°C, an additional intermediate phase is found which has a $(5\sqrt{3}\times 5\sqrt{3})R30^\circ$ periodicity. The corresponding LEED and RHEED patterns are presented in the SM in Fig. S2 \cite{supp}.
The $(\sqrt{3}\times \sqrt{3})R30^\circ$ superstructure (iii and iv) is most likely a Te adlayer, relatively weakly bound to the Te-terminated $\alpha$-MnTe surface atoms and thus stable only up to $250\,$°C.

It remains unsolved why we observe the $(2\times 2)$ superstructure on the surface of all grown $\alpha$-MnTe(0001) films except for the \FGT~case. Neither post-growth annealing nor Te deposition led to the formation of a $(2\times 2)$ reconstruction in the MnTe\,/\,\FGT~system.
To better understand the discrepancy, a detailed knowledge of its exact atomic structure is required. We suggest a quantitative LEED-$I(V)$ structure analysis for this purpose, which, however, goes beyond the scope of this study.
One reasonable explanation is the improved surface morphology of the films grown on \FGT, which we infer from the sharpness of the LEED spots (Fig.~\ref{fig:1}(d-f)). That is, larger terraces on the $\alpha$-MnTe$(0001)$ surface may easier allow Te atoms on the surface to diffuse and occupy the remaining Te vacancies. This further supports the model of a surface-Te-deficient origin of the $(2\times 2)$ superstructure.

In sum, we have shown in this section that the surface atomic structure of the epitaxial $\alpha$-MnTe(0001) films can be systematically controlled via Te deposition and post-growth annealing.

\subsection{Soft X-ray ARPES measurements on $\alpha$-MnTe films}
\label{Sect_ARPES}

\begin{figure*}[t!]
    \centering
    \includegraphics[width=0.9\linewidth]{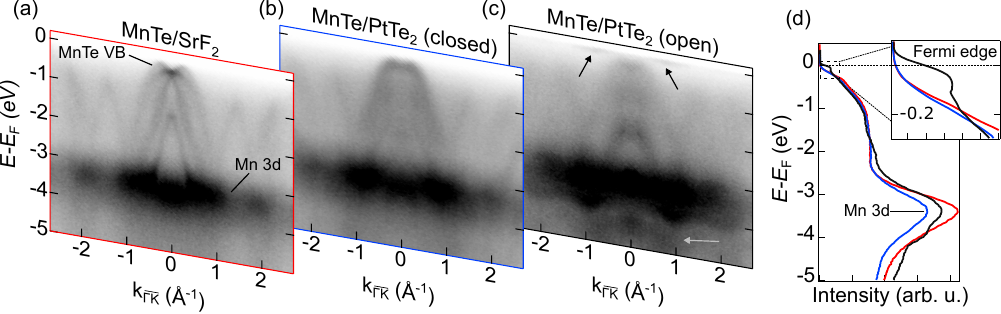}
    \caption{a) Soft X-ray ARPES measurements of $\alpha$-MnTe thin films grown on SrF$_2$ (a) and PtTe$_2$ (b,c) and the corresponding angle-integrated energy distribution curves (EDC) (d).
    The spectra in (a),(b), and (c) have been measured at photon energies of $h\nu = 330\,\mathrm{eV}$, $h\nu = 476\,\mathrm{eV}$, and $h\nu = 326\,\mathrm{eV}$, respectively. All cuts correspond to the $\Gamma$-$\mathrm{K}$ line within the 3D bulk BZ of $\alpha$-MnTe. Here, for each sample, the measurement with the best statistics is shown. Additional data at different photon energies can be found in the SM \cite{supp}.
    The color of the EDC in (d) refers to the respective frame of the ARPES measurement. For PtTe$_2$, two different films are shown, one of which covers the whole substrate (\textit{closed}, (b)), while the other leaves part of the interface exposed (\textit{open}, (c)). In the latter, possible interface features are marked with arrows. The datasets in (a)-(c) are symmetrized to improve statistics and eliminate linear dichroism effects. The spectra in (d) are normalized on the VB intensity. Note that the different intensity of the Mn-$3d$ main peak likely results from different photon energies used.
    }
    \label{fig:ARPES}
\end{figure*}

In the final paragraph of this section, we briefly shed light on the electronic structure of the grown MnTe films. Therefore, we have conducted bulk-sensitive soft X-ray (SX) ARPES measurements, depicted in Fig.~\ref{fig:ARPES}. More comprehensive data, including out-of-plane momentum mapping as well as different high-symmetry cuts, are shown in the Supplementary Material. The measurements presented here have been performed at photon energies corresponding to $\Gamma \mathrm{K} \mathrm{M}$-momentum plane of the bulk Brillouin zone of $\alpha$-MnTe and the $k_\parallel$-path corresponds to a $\Gamma \mathrm{K}$ plane.
The broad, high-intense feature between 3...4\,eV binding energy is attributed to Mn-3d states, whereas the dispersive features correspond to the $\alpha$-MnTe valence bands (VB).
In particular, the films grown on SrF$_2$ (111) (Fig.~\ref{fig:ARPES}(a)) yield sharp bands, reflecting the good crystal quality, with the band splitting at the valence band maximum --- that emerges from the interplay of altermagnetism and spin-orbit coupling \cite{Krempasky2024} --- being clearly visible in our data. The spectrum of the closed film on PtTe$_2$ (Fig.~\ref{fig:ARPES}(b)) yields essentially the same band dispersion, with slightly broader linewidths, as can be expected from a poorer film quality (compare LEED in Fig.~\ref{fig:1}(e)).

Note that, although all spectra were taken on the $(2 \times 2)$-reconstructed samples, we do not find any influence of this reconstruction on the electronic (bulk) band structure. This is not expected for the bulk states observed here in our SX-ARPES measurements but could be relevant for the investigation of surface states \cite{zhao2026,zhou2026}. In fact, $\alpha$-MnTe$(0001)$ surfaces are predicted to host metallic surface states \cite{zhao2026,zhou2026}. Notably, we do not observe a Fermi edge in our measurements, neither in films grown on insulating SrF$_2$ (Fig.~\ref{fig:ARPES}(a)) nor in closed films on metallic PtTe$_2$ substrates (Fig.~\ref{fig:ARPES}(b)). This can also be seen in the corresponding angle-integrated spectra depicted in Fig.~\ref{fig:ARPES}(d); both (red and blue) spectra do not show a Fermi edge but a valence band maximum at $\approx 40\,\mathrm{meV}$ below $E_\mathrm{F}$, which is in line with the expectation for a $p$-doped semiconducting $\alpha$-MnTe film \cite{Kriegner2016}. Also, the metallic PtTe$_2$ substrate does not cause a Fermi edge, given that the film thickness $\gg 50\,\mathrm{nm}$ strongly exceeds the photoemission probing depth. This proves that this film (Fig.~\ref{fig:ARPES}(b)) is completely closed and no substrate areas are exposed to the surface, in line with the observed streaky RHEED pattern (Fig.~\ref{fig:1}(e)).
Another MnTe film grown on PtTe$_2$ is shown in Fig.~\ref{fig:ARPES}(c) (black line in Fig.~\ref{fig:ARPES}(d)). Although this film has also crystallized in the desired hexagonal $\alpha$-MnTe phase, it has a significantly worse film quality. The RHEED pattern (Fig.~S5 in the SM \cite{supp}) has dotted integer streaks, indicating scattering at 3D islands and thus a rough film. In the SX-ARPES measurement on this sample (Fig.~\ref{fig:ARPES}(c)), the same $\alpha$-MnTe features as in the previous cases can be found, just with less intensity as compared to high-quality films. In addition, one can recognize states (highlighted by the black and grey arrows) that are absent in the spectra of the closed films. In particular, there are metallic states right at the chemical potential, and one finds a clear Fermi edge, as is best seen in the angle-integrated spectra in Fig.~\ref{fig:ARPES}(d). This shows that if there is dominant island growth, as expected from the RHEED pattern of this film, the hole defects in the MnTe film penetrate down to the substrate. As such, expose a sizable substrate area, which here gives rise to a metallic PtTe$_2$ photoemission signal. This is complementary supported by the clear Pt-$4f$ signal observed in the core level spectra (Fig.~S5 in the SM \cite{supp}). Overall, the spectrum in Fig.~\ref{fig:ARPES}(c) demonstrates substrate metallicity at the interface after film growth.

\section{Conclusions}
\label{Sect_Discussion}

We have established the epitaxial growth of $\alpha$-MnTe exceeding conventionally used substrates, such as SrF$_2$, and extending the range to different vdW materials. Our study demonstrates the possibility of growing $\alpha$-MnTe on vdW substrates with large lattice mismatches ranging from $-5.3\,\%$ to $4.0\,\%$. For these, the observed XRD peaks (see Fig.~\ref{fig:1}(c)) are at the same position, indicating the identical lattice constants and thus relaxed film growth, which is supported by the vdW-type interfaces.
The substrates used have different physical properties. In particular, $\alpha$-MnTe on the transition-metal dichalcogenide PtTe$_2$ could be of interest to future studies. The large spin Hall conductivities in PtTe$_2$, for example, may promote spin–orbit torque generation across the interface to MnTe. Moreover, potential interface effects arising from the proximity of the altermagnet to the ferromagnetic moments in \FGT~are worth exploring in the future.
Next to the variation of the substrates, we demonstrate the possibility to modify the surface atomic structure. A series of superstructures has been found which can be formed reversibly by Te evaporation and post-growth annealing. Although the specific atomic arrangement of these surfaces remains unknown and requires a quantitative structure analysis, we here demonstrate the existence of these structures and the possibility of modifying them in a controlled manner. Both are extremely relevant to further investigations of the surface electronic structure of altermagnetic $\alpha$-MnTe$(0001)$.
The bulk band structure, investigated here, is not affected by the observed $(2\times 2)$ superstructure.

\section{Acknowledgments}

This work has been supported by the Deutsche Forschungsgemeinschaft (DFG) through the Würzburg-Dresden Cluster of Excellence \textit{ctd.qmat} (EXC 2147, Project ID 390858490) and SFB1170 'ToCoTronics' (Projects A01 and B07) and DFG Project RE1469-13-2.
We acknowledge DESY (Hamburg, Germany), a member of the Helmholtz Association HGF, for providing experimental facilities. Parts of this research were carried out at PETRA III using beamline P04. Beamtime was allocated for proposals I-20241147 and II-20250821.
The ASPHERE III photoemission spectroscopy instrument at beamline P04 was funded by the German Federal Ministry of Education and Research (BMBF) and the German Federal Ministry of Research, Technology, and Space (BMFTR) through the ErUM framework program (projects 05KS7FK2, 05K10FK1, 05K12FK1, 05K13FK1, 05K19FK4, and 05K25FK4 with Kiel University, and projects 05KS7WW1, 05K10WW2, 05K19WW2, and 05K25WW1 with the University of Würzburg)

\bibliographystyle{apsrev4-2}
\bibliography{MnTe}

\end{document}